# High-Efficiency Ultra-Violet Dielectric Meta-Holograms with Antiferromagnetic Resonances

Kun Huang[1,2,†], Jie Deng[2,†], Hai Sheng Leong[2,†], Sherry Lee Koon Yap[2], Ren Bin Yang[2], Jinghua Teng[2], Hong Liu[2]*

**Affiliations:**

[1]Department of Optics and Optical Engineering, University of Science and Technology of China, Hefei, Anhui, China 230026

[2]Institute of Materials Research and Engineering, Agency for Science, Technology and Research (A*STAR), 2 Fusionopolis Way, #08-03, Innovis, Singapore 138634

† These authors contributed equally to this work.

*To whom correspondence should be addressed. E-mail: h-liu@imre.a-star.edu.sg

**Metasurfaces with spatially varying subwavelength structures enable full control of electromagnetic waves over a wide spectrum. High-efficiency metasurfaces, especially in a transmission mode, are of practical significance in optical elements and systems, hitherto their operating frequencies have been expanded down to visible-wavelength ranges. Challenges of developing shorter-wavelength metasurfaces originate from electromagnetic loss caused by strong absorption for most high-refractive-index materials. Here we introduce a large-bandgap semiconductor material—niobium pentoxide ($Nb_2O_5$)—to engineer a ultraviolet meta-hologram with a total efficiency of ~81% at 355nm wavelength. This meta-hologram modulates the geometric phase of transmitted circular-polarization light via orientation-varying high-aspect-ratio nano-bricks that are elaborately designed to excite antiferromagnetic resonances. We reveal that the induced antiferromagnetic modes maintain one component (e.g., $E_x$-component) of incident light through even-numbered antiparallel magnetic dipoles (AMDs) but reverse the other component (e.g., $E_y$-component) via odd-numbered AMDs, thus realizing high-efficiency polarization conversion. Our approach might open the door towards high-performance ultraviolet-light nanophotonics and meta-optics.**

As subwavelength-pixel diffractive optical elements, metasurfaces could tailor the local phase and amplitude of reflected or transmitted light by tuning the geometry of structures, such as dimension, shape and orientation [1-6]. Metasurface-based devices including metalenses[7-10], meta-holograms[11-13] and meta-gratings[14] are not only able to realize the functionalities of their traditional counterparts, but also possess some unprecedented properties such as polarization sensitivity[15], non-linear effects[16,17] and broadband operation[18]. More attractively, their ultrathin features make metasurfaces highly desired in integrated electronic and photonic circuits, portal and mobile devices [19].

High-efficiency manipulation of light is the cornerstone driving metasurfaces towards industrial applications and many efforts have been made to enhance their efficiency. In principle, if metal-based reflective metasurfaces are well-fabricated, the optimized efficiency could be as high as ~80% in visible and near-infrared ranges [13,20]. However, the transmission mode is much preferred in most devices and systems, where lossy metallic metasurfaces with plasmonic resonances have low transmission efficiency [12,21]. In contrast, low-loss dielectric metasurfaces employing magnetic resonances [14,22] are the most promising candidates to realize high-efficiency transmission meta-devices. For example, silicon metasurfaces have the efficiency of ~80%-99% in near-infrared range [12,23], while high-efficiency metasurfaces at visible wavelengths have been demonstrated by using other semiconductor materials such as titanium oxide ($TiO_2$) [7], and gallium nitride (GaN) [24]. Hitherto, ultra-violet metasurfaces still remain a virgin land yet exploited, as they pose rigorous requirements on material properties such as large bandgap, high refractive index, and manufacturability of nano-patterning in ultra-small pixel pitches[25]. Although photonsieve holograms[26] perforated in metal films have been demonstrated at the ultra-violet wavelengths [18], their total efficiency ( <1%) is too low for practical applications. Accomplishment of developing high-efficiency ultraviolet metasurfaces would break the bottleneck of nano-photonics for miniaturizing shorter-wavelength devices and systems.

For dielectric metasurfaces, the electric- and magnetic-dipole resonances fundamentally govern scattering properties of nanostructures [27]. Two overlapping electric and magnetic dipole resonances with equal strength lead to extremely high transmission in Huygens metasurfaces [28]. It has been found that the magnetic multipole resonances [21] dominate in geometric metasurfaces that modulate the phase of circularly polarized (CP) transmitted light with its spin opposite to the incident one, through orientation-varying dielectric nanofins. Although these nanofins usually function as nanoscale half-waveplates (HWPs) [29], their underlying mechanisms of reversing one component of incident CP light while maintaining the other, which is required in high-efficiency polarization conversion, have not been investigated comprehensively. A deep insight into the hidden physics might provide guidance for designing high-efficiency metasurfaces.

To address these challenges, we introduce a semiconductor of niobium pentoxide ($Nb_2O_5$) as a new material platform to develop dieletric geometric metasurfaces at near-ultraviolet wavlengths. Due to the large bandgap and high refractive index, a conversion efficiency of ~83% is predicted to convert the incident circular-polarization light to the transmitted light with opposite spins via a $Nb_2O_5$ nano-brick array devised upon geometric metasurfaces operating at the wavelength of 355nm. As an experimental proof, a transmissive meta-hologram has been demonstrated with a total efficiency as high as ~81% and a large angle-of-view of 64°. We find that, both antiferromagnetic modes with even- and odd-numbered AMDs are induced by *x*- and *y*-components of incident circular-polarization light respectively. The even-numbered AMDs can maintain the electric vectors of light at the input and output planes of nanobricks, while odd-numbered AMDs inverse both electrical vectors, thus realizing the polarization conversion between two spins.

**Optical properties of $Nb_2O_5$**

$Nb_2O_5$ has a bandgap of ~3.65 eV (see Fig. 1a), which is ~0.45 eV and 2.55 eV higher than $TiO_2$ and silicon, respectively. Such a large bandgap determines that $Nb_2O_5$ exhibits a wide transparency window spanning ultraviolet, visible and infrared regimes [30]. In addition, its intriguing optical properties of high permittivity and low absorption have enabled a great variety of applications such as solar cells [31] and light emitting diodes [32]. Moreover, $Nb_2O_5$ films can be manufactured by general deposition techniques and synthesized into various shapes such as nanobelts, nanopore, nanorods and nanospheres [30]. All these properties make $Nb_2O_5$ a competent material platform for developing ultraviolet nano-photonics.

In this work, we employ atomic layer deposition (ALD) method to prepare amorphous $Nb_2O_5$ at a relatively lower temperature of 90°C comparing with previous study[33]. It is purposely chosen below the glass-liquid transition temperature ($T_g$, ~105°C) of common electron beam lithography resist, e.g., PMMA or ZEP520A, thereby ensuring high fidelity of pattern transfer. Superior to other coating techniques such as evaporation and sputtering, ALD yields excellent conformal coverage, monolayer control of thickness, ultrasmooth surface and high uniformity, which are the key to acquire high-aspect-ratio nanostructures. To evaluate the optical properties of as-grown $Nb_2O_5$ film, we measured its complex refraction index by using ellipsometer. Figure 1b plots the refraction index of 2.2 (Re($n$)) and almost negligible extinction ratio (Im($n$)) at the ultraviolet wavelengths ranging from 350nm to 400nm, which implies its low loss and high index.

**Design of dielectric metasurfaces**

Figure 2a shows the sketch of one unit cell in geometric metasurfaces composed of orientation-rotating $Nb_2O_5$ nanobricks on quartz substrate. Upon a left-handed CP light of incidence, one part of its transmitted light has the right-handed circular polarization and simultaneously obtains an additional phase that is twice of the rotation angle $\theta$ of nanobricks. The conversion efficiency ($\eta$) from the incident polarization to its cross-polarization is determined by the geometry, *i.e.*, height ($H$), width ($W$) and length ($L$) when its material property, substrate and ambient medium are fixed. In our configuration, we theoretically predict the maximum conversion efficiency of ~83% with an optimized geometry of $H$=430nm, $L$=150 and $W$=70nm, which is simulated at the wavelength of 355nm by using finite-difference time-domain (FDTD) method (see details in Supplementary Note 1). The conversion efficiency is evaluated as the ratio of the power of transmitted light with cross polarization to the total incident power.

**Antiferromagnetic resonances**

To unveil the role of magnetic resonances during polarization conversion, we investigate both electric and magnetic fields resonating with the optimal geometry. Figures 2b and 2c show the electromagntic responses induced by $E_x$ and $E_y$ components of the incident CP light, respectively. For $E_x$ component, its induced electric fields contain four circle displacement currents with alternately clockwise and anti-clockwise directions, which are relative to four antiparallel magnetic dipoles (AMDs) that are vertically located along $z$ axis. We note that such a staggered magnetization behaves like one-dimensional antiferromagnetic chain, which has also been observed recently in other artificial structures such as plasmonic nano-disk array [34] and hybrid metamaterials [35,36]. In such antiferromagnetic modes, the even-numbered circle displacement currents with alternate spins orientate both electric vectors of light at the incident (bottom) and output (top) planes of nanobrick pointing towards the same direction (see Fig. 2b). It implies that the *x*-component of incident CP light is maintained after passing through nanobricks. However, for $E_y$ component, an antiferromagnetic mode with three AMDs is induced, so that these odd-numbered circle currents make the electric vector of transmitted light inverse to that of incident light (Fig. 2c). Thus, the transmitted light has the polarization completely orthogonal to the incident light, manifesting a high-efficiency polarization conversion. Such optically induced anti-ferromagnetic modes in nanoscale HWPs are fundamentally distinguishable from the birefringence effect in traditional bulky HWPs and the electric-dipole resonances in plasmonic nano-HWPs.

Two coupling effects dominate the interaction between the antiferromagnetic modes and the nanobricks. The first coupling comes from the fact that, the dimension of nanobricks determines the number of magnetic dipoles in both antiferromagnetic modes induced by *x* and *y* components of incident light. The long axis (*x*-axis in Fig. 2c) of nanobricks enables these *x*-orientated magnetic dipoles to oscillate in a large volume.. In contrast, a magnetic dipole oscillating along *y* axis (Fig. 2b) has a relatively small volume, which is constrained by the width of 70nm. Since the volume of nanobricks is fixed, the number of *x*-orientated magnetic dipoles (Fig. 2c) is 1 smaller than that of *y*-orientated magnetic dipoles (Fig. 2b). Consequently, even- and odd-numbered AMDs, oscillating along *x* and *y* axes respectively, are excited simultaneously for polarization conversion. Secondly, the magnetic dipole coupling in the same antiferromagnetic mode influences the weight of every dipole in the total magnetic field. A dipole-coupled method (see Methods) is used to calculate this coupling weight. By using the AMDs that are induced by simplified single-ring displacement currents (see Supplementary Note 2), we mimic the antiferromagnetic modes (Figs. 2d and 2e), where every dipole mode has different amplitude and phase by taking account of the coupling effect. As expected, the even- (odd-) numbered dipole modes maintain (reverse) the electric vector of the incident light.

**Ultra-violet meta-hologram**

A meta-hologram is proposed to validate the efficiency of such $Nb_2O_5$ metasurfaces at 355nm wavelength. As sketched in Fig. 3a, this meta-hologram can convert the incident left-handed CP light into the right-handed transmitted light with an additional position-dependent phase profile, and achieve a holographic image at the Fresnel region of 300μm. A modified Gerchberg-Saxton algorithm [21] is used to design the required phase (see Fig. 3b) for reconstructing a holographic image of "rabbit" pattern as shown in Fig. 3c. To reduce diffraction loss, the phase is discretized into 16 levels (with the phase value $\varphi_n=2\pi n/16$ at level *n*), which physically corresponds to the nanobricks with a rotation angle of $\varphi_n/2$.

The hologram is fabricated by using electron-beam lithography and ALD technique (see Supplementary Note 3). Figure 2d shows its scanning-electron-microscope (SEM) images with smooth and nearly-vertical side walls, indicating a precisely patterned device. More SEM images with larger field of view could be found in Supplementary Fig. 4. The sample is measured in a confocal microscope (WITec Alpha 300S) with high-precision stages (see Supplementary Note 4). To increase the incident power on hologram, a weakly focused CP light is used to illuminate the sample, and its transmitted light is collected by an objective lens (Nikon CP Plan, 20×) and then recorded directly by a charge-coupled-device camera (JAI, CM-140MCL-UV), without using polarization analyzer. Figure 3e displays the captured holographic image with a clear contour, high signal-to-noise ratio, the immunity to twin image and high-order diffraction. A quantitative comparison between the experimental and simulated line-scanning intensity profiles is shown in Fig. 3f, where their excellent coincidence confirms high quality of our design and fabrication.

To determine its total efficiency, we take the light shining within the hologram area of 100μm×100μm as the total incidence ($I_{in}$). The output ($I_{out}$) is calculated by using the power encircled only in the "rabbit" pattern without any background. A detailed introduction on efficiency measurement can be referred to Supplementary Note 5. Our experimental efficiency $I_{in}/I_{out}$ is measured as high as 81.6%, which agrees well with the simulated efficiency of ~83%. It confirms that our $Nb_2O_5$ metasurfaces for ultraviolet light could reach comparable efficiency to those meta-devices at visible and infrared wavelengths [12,13,23].

To test its properties, the conversion efficiency and image quality of this hologram are investigated under different oblique incidences. By using the same setup, the oblique incidence is realized by tilting the hologram sample that is fixed onto a manual rotation mount for recording the angle α (see Fig. 4a). Both simulated and measured results in Fig. 4b have good agreement, which benefits from the accurate collection of incident and output power in our experimental strategy. The efficiency decreases with the increment of the tilting angle, which originates from the reduction of z-component (responsible for exciting the antiferromagnetic modes) of wave vector. Similarly, the uniformity and shape of holographic images deteriorate upon increasing the tilting angle, as observed from the experimental and simulated images in Supplementary Fig. 6. To evaluate the image fidelity, the root-means-square errors (RMSEs) of the images taken under oblique and normal illuminations are calculated and plotted in Fig. 4c, where both RMSEs agree well. It shows that image fidelity deteriorates when RMSE increases with the incidence angle α. We also note that, the pattern at the tilting angle 32° is still distinguishable, which implies a larger angle-of-view of 64°×64° compared with previous reports [18].

**Discussion**

As the first demonstration of high-efficiency ultraviolet metasurfaces, the significance of our work is four-fold. Firstly, $Nb_2O_5$ as a semiconductor material has an intrinsic advantage to develop nano-structured devices that can be directly integrated into electronic and photonic devices with the help of matured nanofabrication technology. Our results confirm the feasibility of developing ultraviolet nanophotonics by using nano-structured $Nb_2O_5$ arrays. Secondly, $Nb_2O_5$ exhibits a broader transparency window in the electromagnetic spectrum than $TiO_2$ and Si, thus allowing the development of planar nano-devices beyond the ultraviolet spectrum such as visible as well as infrared wavelengths. Thirdly, as a functionalized device with high efficiency and large viewing angle, our transmissive meta-hologram might accelerate the miniaturization and planarization of traditional ultraviolet elements, devices and systems with bulky volume, thereby benefiting the entire ultraviolet-related industry. Finally, differing from the birefringent effects in traditional HWPs and electric dipole response in plasmonic nano-HWPs[37], the antiferromagnetic modes with AMDs provide a completely new insight into the fundamentals of dielectric nano-HWPs, which thus facilitate a better understanding of the roles of magnetic resonances in tailoring optical properties of subwavelength structures.

In summary, we have demonstrated a $Nb_2O_5$-based dielectric metasurfaces hologram with a total efficiency of ~81%, high-quality image and large angle of view at the wavelength of 355nm. We reveal that the optically induced antiferromagnetic modes hold the underlying physics of governing the polarization conversion in a dielectric nano-HWP, which offers a new paradigm of employing nanoscale magnetic modes to elucidate new phenomena in nano-structures. The proposed $Nb_2O_5$ enriches the choices of materials in developing ultra-violet nanophotonics and flat-optics with advanced nanofabrication techniques, and therefore might spur a wide spectrum of applications such as optical nanolithography, biology, anti-counterfeiting and imaging.

**Methods**
**Dipole-coupled method.** To determine the total electromagnetic fields near the displacement currents, the dimension of these optically induced magnetic dipoles in Figs. 2b and 2c cannot be ignored. The frequently used magnetic dipoles with the far-zone approximation [38], where the dipole volume is taken to be infinitesimal, are not suitable for our case. Here, we employ the magnetic dipoles induced by a theoretical single-ring displacement current, the fields of which can be found in Supplementary Note 2.

A linearly polarized beam propagating along $z$ direction has its electric and magnetic fields of $\mathbf{E}_{in}=E_0 \cdot e^{ikz}\mathbf{e}_x$ and $\mathbf{H}_{in}=(\varepsilon/\mu)^{1/2} E_0 \cdot e^{ikz}\mathbf{e}_y$ respectively, where $k$ is the wavenumber of light, $\varepsilon$ and $\mu$ are the permittivity and permeability of ambient medium, respectively. As observed in Figs. 2b and 2c, the electric dipole resonances in the nanobrick are so weak that they are ignored in our simulations. In addition, these optically induced ring currents are tangent to each other, so that two neighboring single-ring currents in our simulations are also set with the same condition by sharing the only tangent point. It leads to the weak coupling effects of electric fields, which are neglected here. Therefore, we only consider the coupling between these magnetic dipoles.

In order to take the similar form as the previous dipole-couple method [38,39], the coupling effect is calculated by using the magnetic fields located at the original point of each single-ring current. We assume that, for the $i$-th single-ring current, the magnetic field at its center point (with its position labelled by $\vec{r}_i$) of the ring is taken as $\mathbf{H}_i(\vec{\mathbf{r}}_i)$. According to the dipole-coupled theory [40], the magnetic field $\mathbf{H}_i(\vec{\mathbf{r}}_i)$ is equal to a coherent superposition of magnetic fields of incident light and other dipoles. It can be mathematically expressed as

$$\mathbf{H}_i(\vec{\mathbf{r}}_i) = \mathbf{H}_{in}(\vec{\mathbf{r}}_i) + \sum_{j \neq i} \mathbf{H}_j(\vec{\mathbf{r}}_j - \vec{\mathbf{r}}_i), \quad (1)$$

where $\mathbf{H}_j$ is the magnetic field of *j*-th dipole with its center point of ring located at the position of $\vec{\mathbf{r}}_j$, and $\vec{\mathbf{r}}_i$ denotes the position of the center point for the *i*-th single-ring current. In Eq. (1), $\mathbf{H}_{in}(\vec{\mathbf{r}}_i)$ and $\mathbf{H}_j(\vec{\mathbf{r}}_j - \vec{\mathbf{r}}_i)$ are the magnetic fields of incident light and *j*-th dipole at the postion $\vec{\mathbf{r}}_i$, respectively. For the *i*-th magnetic dipole, its magnetic field $\mathbf{H}_i$ has two unknown variables of the coupling coefficient $I_i$ and the ring radius of *a*, as seen in Supplemenatry Note 2. In our simulations, the ring radius is approximately set to be *a*=*H*/(2*N*), where the height (*H*) of the optmized nanobrick is 430nm and *N* is the number of magnetic dipoles confined in nanobricks. Thus, Eq. (1) forms a group of *N* linear equations that contain *N* variables of $I_i$ (*i*=1, 2, ..., *N*), which can be solved directly by using general linear algebra. Hence, the total magnetic fields can be obtained by using a superposition of all the fields of magnetic dipoles, *i.e.*, $\mathbf{H} = \sum_{i=1}^{N} \mathbf{H}_i(\vec{\mathbf{r}})$, as shown in Figs. 2d and 2e.

**Hologram Design.** The proposed meta-hologram is a Fresnel hologram that has its imaging plane located at the meso-field of 300μm, as shown in Fig. 3a. An accurate simulation of its propagation in free space can be implemented by using Rayleigh-Sommerfeld diffraction integral, which is realized by using our developed method based on fast-Fourier-transform (FFT)[41]. In principle, the angular-spectrum method can be also used to simulate the propagation of light, where evanescent frequencies contained in the metasurface-modulated optical fields must be dropped numerically in the frequency domain. This numerical operation causes that the simulated fields might have some suddenly changed phase, which will influence the phase retrieval in the reconstructing a holographic image via multiple iterations. Therefore, the angular-spectrum method is usually used in the cases of single propagation without multiple iterations, such as simulating the optical vortex fields[42]. In contrary, our FFT-based Rayleigh-Sommerfeld diffraction method has no such issue, which can be observed and validated from our simulated and experimental results.

**Acknowledgement**

The authors would like to thank S. L Teo and Doreen M. Y. Lai for the help on taking SEM images. The authors acknowledge financial support from Agency for Science, Technology and Research (A*STAR) Grant (no. 162 15 00025). K. H. thanks the support from CAS Poineer Hundred Talents Program, "the Fundamental Research Funds for the Central Universities" in China, and the National Natural Science Foundation of China (Grant No. 61705085).

**Author contributions**

K. H. and H. L. conceived the idea. K. H. developed the theory, performed the simulations and designed the holograms. J. D., S. Y., R. Y. and H. L. fabricated and characterized the samples. H. S. L. and K. H. built the experimental setup and implemented the optical measurement. K. H., J. T. and H. L. wrote the manuscript. H. L. and K. H. supervised the overall project. All authors discussed the results, did the data analysis and commented on the manuscript.

**Competing financial interests**

The authors declare no competing financial interests.

**Figures**

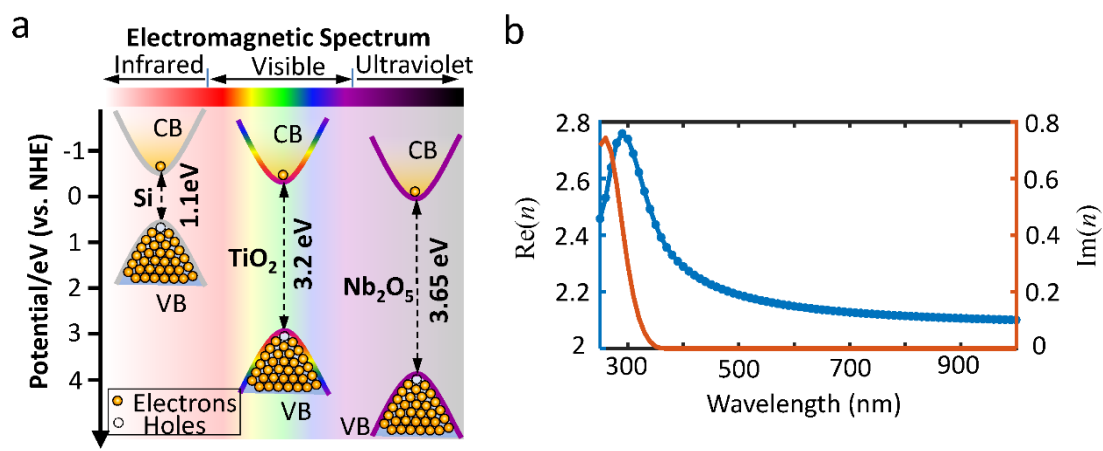

**Figure 1 | Optical properties of $Nb_2O_5$.** (**a**) The bandgap diagram of silicon (Si), titanium dioxide ($TiO_2$) and niobium pentoxide ($Nb_2O_5$) at infrared, visible and ultra-violet spectra respectively. The potential are with respect to normal hydrogen electrode (NHE). The conduction (CB) and valence bands (VB) of electron in materials are sketched by colored curves. (**b**) Measured complex refractive index of our ALD-grown $Nb_2O_5$ film.

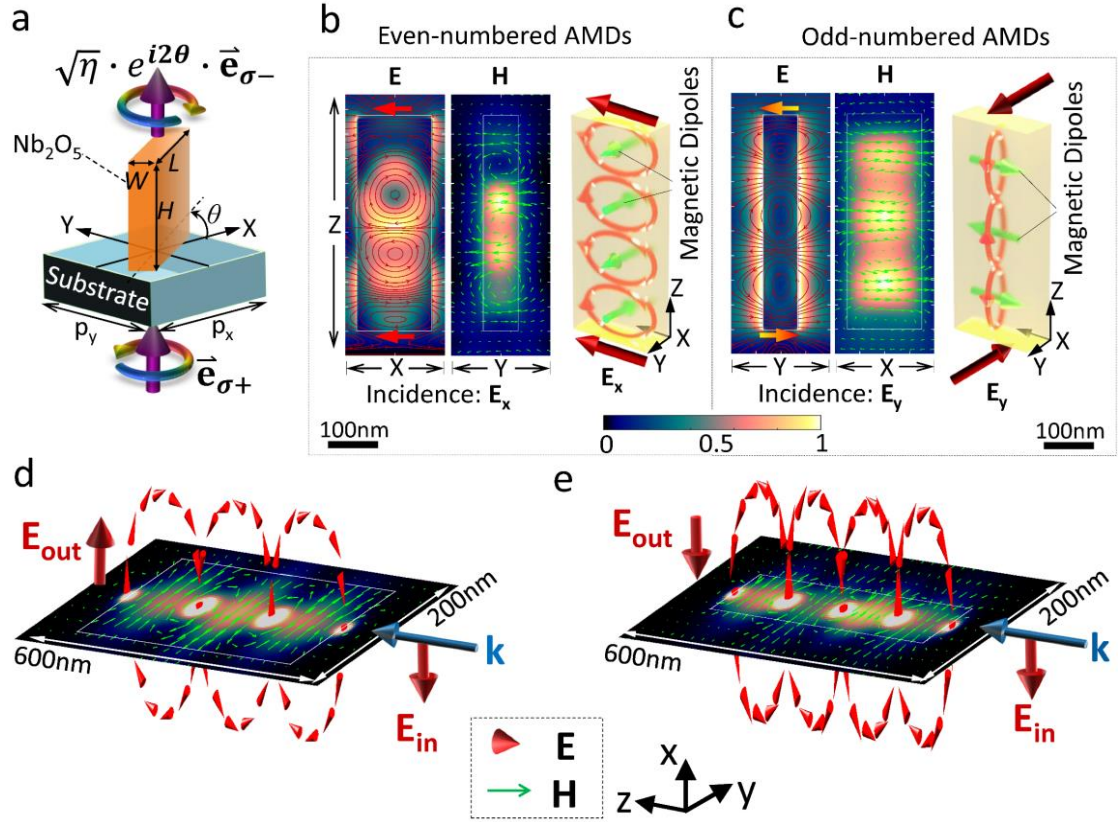

**Figure 2 | Optical antiferromagnetic resonances in geometric metasurfaces. (a)** Sketch of a cell unit with the dimension of $p_x=p_y=250$nm. The $Nb_2O_5$ nanobrick with a rotation angle of $\theta$ is located on a quartz substrate and used to convert the incident circular polarization into the transmitted cross-polarization with the efficiency $\eta$ and an additional phase modulation of $e^{i2\theta}$. **(b-c)** Electromagnetic resonances existing in the optimized nanobrick. The vector profiles of induced electric (**E**) and magnetic fields (**H**) are distinguished by using the red lines and green arrows, respectively. The white rectangles denote the boundaries of nanobrick. The vertically staggered magnetic dipole moments in an anti-parallel alignment, *i.e.,* anti-ferromagnetic modes, are excited and confined within the nanobrick. The anti-ferromagnetic modes with even- **(b)** and odd-numbered **(c)** dipole moments are induced by $E_x$ and $E_y$ components of the incident light, respectively. The inserts show the simplified sketches of two antiferromagnetic modes. **(d-e)** The antiferromagnetic modes composed of the odd **(d)** and even-numbered **(e)** anti-parallel magnetic dipoles that are induced by single-ring currents (taking as the approximation of optically induced volume currents in (**b-c**)). The vectors of single-ring currents are represented by three-dimensional red arrows. The coupling weight of every magnetic dipole is calculated by using the dipole-couple method. To simplify our simulation, all the sing-ring currents have the same radius of 430nm/(2*N*), where *N* is the number of magnetic dipoles.

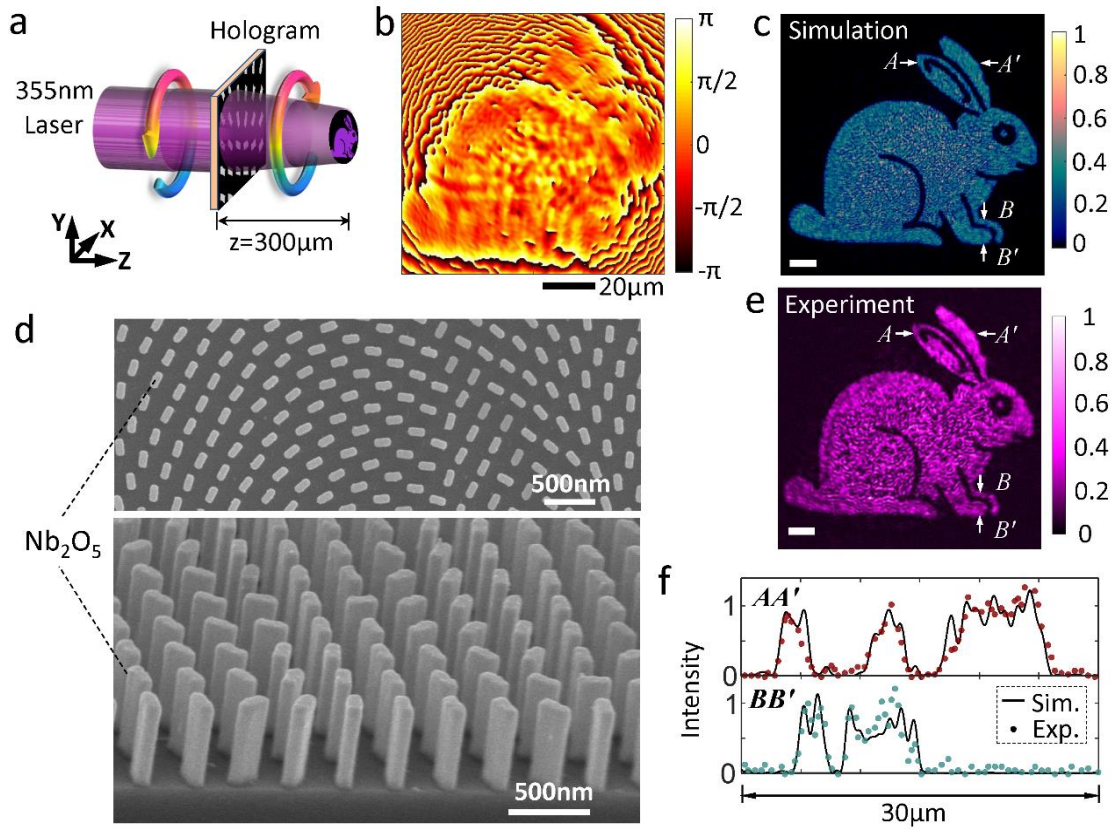

**Figure 3 | Ultraviolet meta-hologram.** (**a**) Working principle of this meta-hologram converting the left-handed circular polarized light into the right-handed one and reconstructing a holographic image at the distance of $z$=300μm. The circular arrows represent the left- (before the hologram) and right-handed (after the hologram) circularly polarized light, respectively. (**b**) The designed phase profile with the size of 100μm×100μm and pixel pitch of 250nm×250nm. The lines $AA'$ and $BB'$ denote the locations of line-scanning intensity profiles for data analysis. (**c**) Simulated holographic image with a pattern of "rabbit". Scale bar: 10μm. (**d**) The SEM images of our fabricated samples with top (upper panel) and tilting (lower panel, 60°) view. Scale bar: 500nm. (**e**) Captured holographic image by CCD camera. Scale bar: 10μm. (**f**) Simulated and measured one-dimensional intensity profiles along the lines with their locations denoted by $AA'$ and $BB'$ in (**c**) and (**e**).

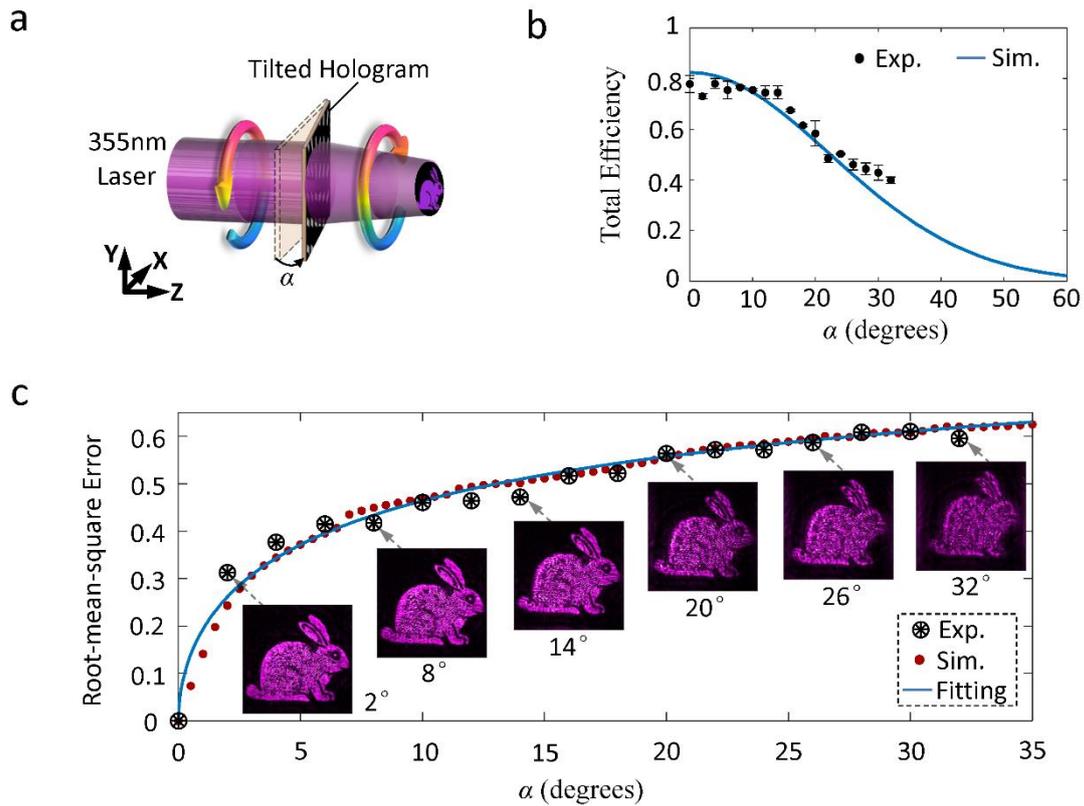

**Figure 4 | Tolerance to the oblique incidence.** (**a**) Sketch of realizing the oblique incidence by tilting the hologram. The other components in the whole system maintains with the same alignment. (**b**) Simulated and measured efficiency at the different incident angles ranging from 0° to 60°. Due to the sufficient phase level of 16, all the modulated cross-polarized light is considered to contribute the holographic image. Therefore, the simulation data (solid curve) obtained directly from FDTD calculations can be taken as the theoretical total efficiency. (**c**) The RMSEs of images between the normal and oblique incidence. Simulated RMSEs are labelled by red dots and fitted by a solid curve while experimental results are shown by encircled asterisks. The inserts show several experimentally captured images taken under the incidence angles with an interval of 6°.

# Supplementary Materials

**Supplementary Note 1 | Design and simulation of $Nb_2O_5$-based geometric metasurfaces**

To evaluate the performance of metasurfaces, the simulation is implemented by using the finite-difference time-domain method in a commercial software of Lumerical. In our simulations, the computational area is set to be 250nm×250nm×2000nm along *x*, *y*, and *z* directions, respectively. To investigate the optical response of periodical nanostructures, the periodic boundary conditions are employed along *x* and *y* directions, while the perfectly match layers are used along *z* directions to eliminate the unwanted reflections from top and bottom of calculation window. The material properties (i.e., complex refractive index) of $Nb_2O_5$ as shown in Fig. 1b are adopted in our simulations.

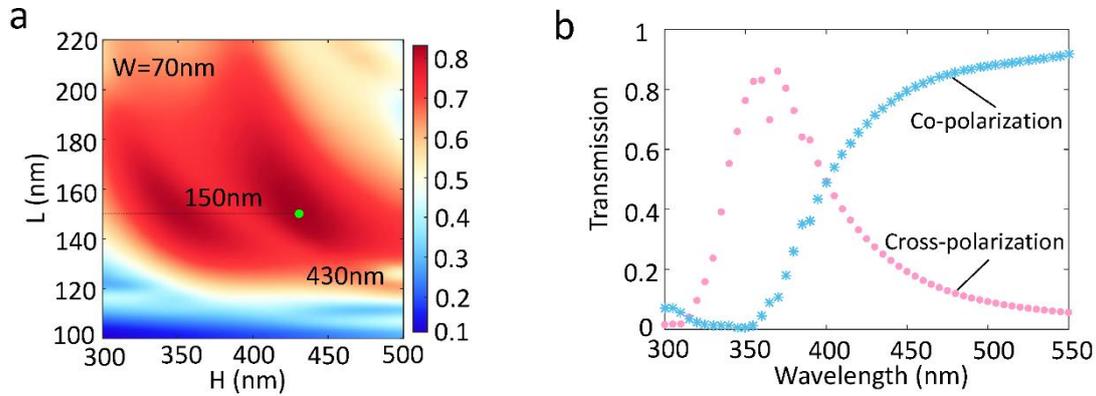

**Supplementary Fig. 1.** (a) Simulated conversion efficiency $\eta$ for nanobricks with a fixed width 70nm, various heights *H* and lengths *L*. The peak efficiency of ∼83% is located at *W*=70nm, *H*=430nm and *L*=150nm. The simulation is carried out in FDTD by using the measured refractive index in Fig. 1b of main text. (b) Broadband response of the optimized nano-brick array. In the transmitted light, the efficiency values are shown by addressing the co-polarization and cross-polarization parts relative to incident polarization. The co-polarization transmitted light has no phase modulation.

The refractive index of quartz substrate is retrieved from Palik's data [S1]. The incident light at the wavelength of 355nm is composed of two orthogonal *x*-polarized and *y*-polarized plane-wave sources with their phase delay of $\pi/2$, thus generating the required circular polarization.

To fully map conversion efficiency over different geometry of nanobricks, a self-made script file is employed to re-call the main program by changing the geometric parameters of nanobricks in every loop. Considering the limitation of fabrication, we set the width of nanobrick to be *W*=70nm, which is achievable by using our EBL. Supplementary Fig. 1a shows the simulated efficiency with the length varying from 100nm to 220nm and the height from 300nm to 500nm. The simulated efficiency is calculated by using the ratio of the transmitted light with cross polarization to the total incident light. It can be seen that the peak efficiency of ~83% is attainable with *L*=150nm and *H*=430nm. In addition, the neighboring geometries also exhibit efficiency of > 80%, thereby allowing a large tolerance to fabrication error.

The broadband response of such optimized nanobrick is also investigated. Supplementary Fig. 1b shows the simulated efficiency at the wavelengths from 300nm to 550nm. The efficiency of transmitted light with cross-polarization is ~80% around the designed wavelengths of 355nm and decreases as the wavelength deviates. The shifted wavelength modifies the magnetic modes existing in the nanobrick so that the antiferromagnetic resonances deteriorate, leading to the diminishing conversion efficiency. For the efficiency larger than 50%, the simulated bandwidth is ~60nm ranging from 340nm to 400nm. In addition, the total transmission (including co-polarization and cross-polarization light) is above 80% when the wavelength is larger than 350nm, which implies that $Nb_2O_5$ nanostructures with optimized geometry could also work as high-efficiency visible-light metasurfaces.

**Supplementary Note 2 | Magnetic dipoles induced by a single-ring current**

As illustrated in Supplementary Fig. 2a, we assume that the optical single-ring displacement current with its radius *a* locates at the *x-y* plane so that we can discuss its induced magnetic dipole in the commonly used polar coordinate, enabling a better understanding. Because the single-ring displacement current is induced by optical fields, we take such single-ring displacement current as a localized oscillating source with a sinusoidal dependence on time. Thus, its current density could be expressed as

$$\mathbf{J}(\mathbf{x},t) = \mathbf{J}(\mathbf{x})e^{-iwt}, \qquad (S1)$$

where *w* is the time frequency of light and $\mathbf{J}(\mathbf{x})$ is the vectorial current density without the time dependence. For the single-ring displacement current, $\mathbf{J}(\mathbf{x})=I\cdot\delta(r-a)\cdot\delta(z)\mathbf{e}_\varphi$, where *r*, *φ* and *z* are the polar coordinates, *I* is a constant that determines the amplitude and phase of the induced magnetic dipole, *a* is the radius of the single-ring displacement current, the delta functions limit the current flow within a ring with its radius of *a* at the cut plane of *z*=0. In the Lorenz gauge, its vector potential $\mathbf{A}(\mathbf{x}')$ at the position of $\mathbf{x}'$ is written as (see Equation 9.3 in Ref. [S2])

$$\mathbf{A}(\mathbf{x}') = \frac{\mu}{4\pi}\iiint \mathbf{J}(\mathbf{x}')\frac{e^{ik|\mathbf{x}'-\mathbf{x}|}}{|\mathbf{x}'-\mathbf{x}|}d^3x, \qquad (S2)$$

where *μ* is the permeability of the ambient medium, *k=w/c* is the wave number and *c* is the speed of light. After a simple substitution, we can find the vector potential

$$\mathbf{A}(\mathbf{x}') = \frac{\mu}{4\pi}\int_0^{2\pi}d\varphi\int_0^\infty rdr\int_{-\infty}^\infty dz\cdot[I\cdot\delta(r-a)\cdot\delta(z)\frac{e^{ikR}}{R}]\begin{bmatrix}-sin\varphi\\cos\varphi\\0\end{bmatrix}$$

$$= \frac{\mu Ia}{4\pi}\int_0^{2\pi}\frac{e^{ikS}}{S}\begin{bmatrix}-sin\varphi\\cos\varphi\\0\end{bmatrix}d\varphi, \qquad (S3)$$

where $R= [(\rho\cos\phi-r\cos\varphi)^2+(\rho\sin\phi-r\sin\varphi)^2+(z'-z)^2]^{1/2}=[r^2+\rho^2-2r\rho\cos(\phi-\varphi)+(z'-z)^2]^{1/2}$, $S=[a^2+\rho^2-2a\rho\cos(\phi-\varphi)+z'^2]^{1/2}$, $\rho$, $\phi$ and $z'$ are the polar coordinates in the space $\mathbf{x'}$ of interested. From Eq. (S3), we find that the vector potential $\mathbf{A}(\mathbf{x'})$ has only the azimuthal component $A_\varphi = \frac{\mu I a}{4\pi}\int_0^{2\pi}\frac{e^{ikS}}{S}\cos(\phi-\varphi)d\varphi$. Thus, we have its magnetic field [S2]

$$\mathbf{H}(\mathbf{x'}) = \frac{1}{\mu}\nabla\times\mathbf{A}(\mathbf{x'}) = -\frac{1}{\mu}\frac{\partial A_\varphi}{\partial z'}\mathbf{e}_\rho + \frac{1}{\mu}\left(\frac{A_\varphi}{\rho}+\frac{\partial A_\varphi}{\partial\rho}\right)\mathbf{e}_{z'}, \quad (S4)$$

and its electric field outside the ring current

$$\mathbf{E}(\mathbf{x'}) = \frac{i}{k}\sqrt{\frac{\mu}{\varepsilon}}\cdot\nabla\times\mathbf{H}(\mathbf{x'}), \quad (S5)$$

where $\varepsilon$ is the permittivity of the ambient medium. Equations S4 and S5 are the accurate expression of the electromagnetic fields for such single-ring displacement current, without any approximation. Therefore, they allow us to investigate the fields near the displacement current by using the numerical integration that is available in many scientific software such as Mathematica and Matlab.

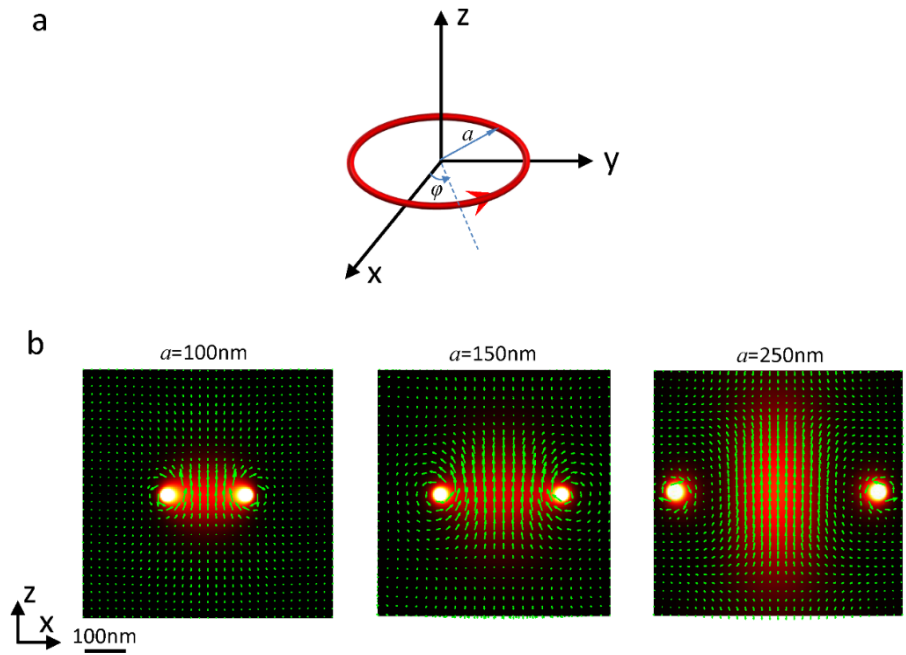

**Supplementary Fig. 2. The electromagnetic fields induced by a single-ring displacement current.** (a) Sketch of a single-ring displacement current in the polar coordinate. (b) Simulated intensity and vector profiles of magnetic fields induced by single-ring displacement currents with the radii from $a$=100nm, 150nm and 250nm.

Supplementary Fig. 2b shows the magnetic fields induced by the single-ring displacement current with different radii $a$=100nm, 150nm and 250nm in the ambient medium of air. In each case, the white area with two bright spots denotes the location of such single-ring displacement current. The magnetic fields in the circle, *i.e.* between two bright spots, show the *z*-orientated dipoles, which can be observed from the vector profiles. The intensity profiles indicate that the volume of the induced magnetic dipoles is large for a large radius $a$, which provides the sufficient space for the magnetic resonances. It determines the number of magnetic dipoles in the nanobricks, as discussed in the main text.

In Eq. (S4), the magnetic fields are determined by the constant $I$, the radius $a$ and the ambient medium (*i.e.*, $Nb_2O_5$). In the dipole-couple theory, the constant $I$ could be adjusted to obey the coupling effect. Considering that the radii of these single-ring displacement currents and their center position are assumed to be fixed in our simulation, the constant $I_i$ is the only variable of the $i$-th magnetic dipole. The calculated coupling coefficients are [-0.4546-0.8351*i*, 0.5385+0.8426*i*, -0.6927-0.6463*i*] for the 3-dipole case in Fig. 2d and [0.5008+0.7145*i*, 0.6491-0.7607*i*, 0.5661+0.6235*i*, 0.5058-0.7920*i*] for the 4-dipole case in Fig. 2e. The magnetic fields in Figs. 2d and 2e have some deviations from those (Figs. 2b and 2c) simulated by using FDTD. The reasons are three-fold. Firstly, the real displacement currents in Figs. 2b and 2c have a fixed volume, which is simplified into a single ring with infinitesimal width in our dipole-coupled method. Secondly, the varying radii and center positions of displacement currents are considered to be fixed in the dipole-coupled method, as mention above. Thirdly, the coupling effect between antiferromagnetic modes and the geometry of nanobricks cannot be taken into account in the dipole-couple method. It indicates that the interaction between antiferromagnetic modes and nano-structures is so complicated that all the issues, such as the volume, size, position and coefficients of dipoles, and the boundary condition between nanostructures and surrounding medium, have to be calculated together. It is difficult to give an analytical solution of the antiferromagnetic modes in such a nano-brick geometry. The magnetic fields in Figs. 2d and 2e are employed here to theoretically validate the existence of antiferromagnetic modes. If one wants to calculate the far-field radiation of such antiferromagnetic modes, the magnetic dipoles can be approximated by using the far-field magnetic dipoles that are commonly adopted in solving various scattering problems [S3-S5].

**Supplementary Note 3 | Fabrication of $Nb_2O_5$ metasurfaces**

The fabrication of our $Nb_2O_5$ metasurfaces is implemented by combining electron-beam lithography (EBL), atomic-layer deposition (ALD) and dry etching techniques. Supplementary Fig. 3 schematically summarizes its fabrication process. It started from spin-coating of a 450 nm thick ZEP resist onto a quartz substrate (15 mm x 15 mm x 0.4 mm) followed by baking at 180°C for 2 minutes. Then Espacer 300Z was spin-coated at 1500 rpm atop the resist. Nano-patterning on the resist was carried out by electron beam lithography ((ELS-7000, Elionix) with an exposure dosage of 360 μC/cm$^2$, acceleration voltage of 100 kV and beam current of 100 pA. After that, the sample was rinsed with DI water to remove the Espacer followed by the development in oxylene for 30 s and then rinsed in IPA for 20 s and blown dry with $N_2$ gas (see Supplementary Fig. 3a). Amorphous $Nb_2O_5$ was grown with an ALD system of Beneq TFS 200 by reacting (tert-butylimino)tris(diethylamino)niobium and water at 90°C (Supplementary Fig. 3b). The niobium precursor was heated up to 70°C in the hot source precursor bottle. Pulse and purge durations for the niobium precursor were 1s and 9s while they were 0.3s and 11s for water, respectively. The growth rate at 90°C was 0.06 nm per cycle measured by ellipsometry. The conformal nature of ALD growth allows the film to grow along the direction normal to the inner wall surfaces of resist. Once the film thickness reaches a minimum value of $t_{resist} = W/2$, the gap will be exactly filled for an ideal case. In practice, a thicker film ($t_{resist} > W/2$) will be formed to avoid any pin-hole at the interfaces of films grown from all the inner surfaces upon completion of ALD process (Supplementary Fig. 3c). Excess $Nb_2O_5$ film (~80nm thickness) was removed by an inductively coupled plasma-reactive ion etching (ICP-RIE) system (OIPT Plasmalab System 100). The ICP RF power was kept constant at 900W and the bias power was set at 150W; The processing pressure was 25 mTorr with CHF3 gas flow of 25 sccm; Helium pressure was 5 Torr for backside-cooling while substrate temperature was maintained at 20°C by a chiller. An etch rate of 48 nm/min was obtain via the measurement of thin film analyzer (Filmetrics F20) (Supplementary Fig. 3d). The residual resist (embedded with $Nb_2O_5$ nanobricks) was removed by RIE etcher (Trion, Sirus T2) with the conditions of chamber vacuum pressure of 250mTorr, power of 250 W, and oxygen ($O_2$) flow of 20 sccm, processing time of 3 minutes (Supplementary Fig. 3e). It can be observed that the height ($H$) of nanobrick is eventually determined by the thickness of resist ($t_{resist}$).

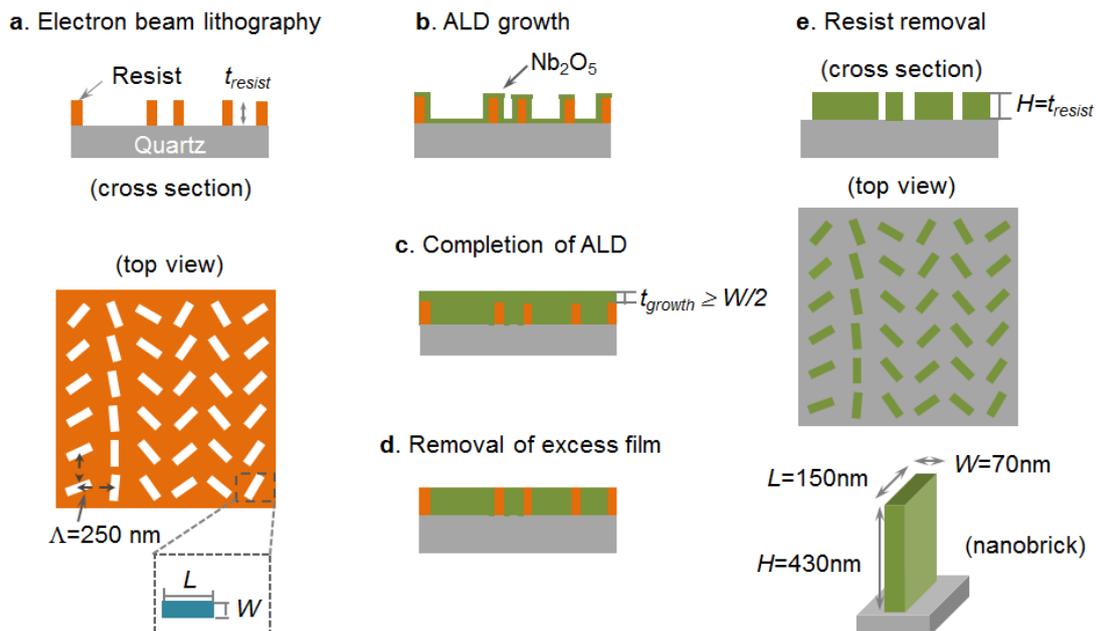

**Supplementary Figure 3. Flow process to realize UV light metasurfaces. (a)**. Electron beam lithography of quartz substrate to transfer inverse pattern of the nanobricks into 450nm thick ($t_{resist}$) ZEP resist (cross section). The period Λ (center-to-center spacing) of the unit cell is about 250 nm. The dashed box shows the close-up view of nanobrick of the length $L$ and width $W$ (top view). **(b)** $Nb_2O_5$ film conformally deposits atop the resist and the sidewalls (cross section). **(c)** Completion of growth with a thickness of film larger than half of the minimum feature ($t_{growth} \geq W/2$) yields a pin-hole free film via ALD. **(d)** Removal of excess film atop resist via an ICP etching step using $CHF_3$ gas. **(e)**. Residual resist removal by $O_2$ plasma etching to realize the $Nb_2O_5$ nanobricks (cross section and top view) with the features of $W = 70$nm, $L = 150$ nm, and $H = 430$ nm.

The SEM images of our fabricated samples are shown with different fields of view in Supplementary Fig. 4. The vertical walls and sharp corners in nanobricks could be seen clearly in Supplementary Figs. 4(a) and 4(b), which imply the excellent fabrication. We have to highlight that our fabricated sample exhibits high uniformity and ultrasmooth surface, as shown by the images with large fields of view in Supplementary Figs. 4(c) and 4(d).

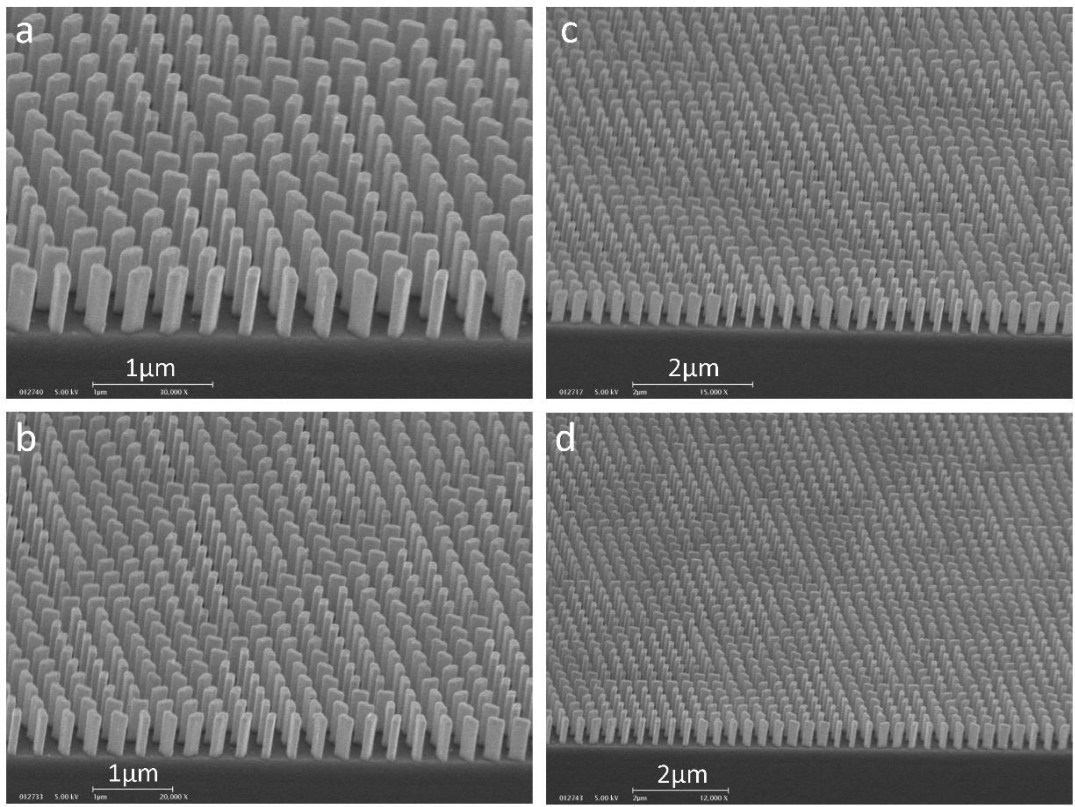

**Supplementary Figure 4.** SEM pictures of our fabricated sample. These images are captured in a view angle of 60 degrees and shown in different fields of view: **(a)** 3.3μm×4.4μm, **(b)** 4.8μm×6.4μm, **(c)** 6.5μm×8.7μm, **(d)** 8.1μm×10.8μm.

**Supplementary Note 4 | Experimental measurement of holographic images**

To facilitate the experimental measurement, we built our setup in a configuration of confocal microscopy (WITec 300S, Germany) that supports the open and self-made optical systems. In this case, we can use the integrated nano-precision stages for flexible tuning and alignment of the fabricated sample, thus allowing for a better characterization. In our experiment (see Supplementary Fig. 5), a laser with its wavelength of 355nm (Cobolt Zouk™) is coupled into a single-mode fiber for obtaining a fundamental Gaussian mode with a good beam profile. The divergent light coming from the single-mode fiber is collected by a thin lens (with its focal length of 35mm) coated with UV-enhanced films, which is used to reduce the loss of light incident on the sample. Due to the limited power (~6 mW measured from the output port of the fiber) of our laser, the collection thin lens is tuned by a 3-axis high-precision stage so that the incident light is weakly focused onto the sample. Thus, the power incident on the sample is increased comparing with the collimated illumination. Here, we have to clarify that such a weakly focused light beam can be taken as a paraxial beam, because the distance between this thin lens and our sample is ~1100mm, which is much larger than the distance (~40mm) between the lens and optical fiber. Therefore, such the weakly focused incident light will not influence our experimental result, as observed in Figs. 3e, 4b and 4c of main text.

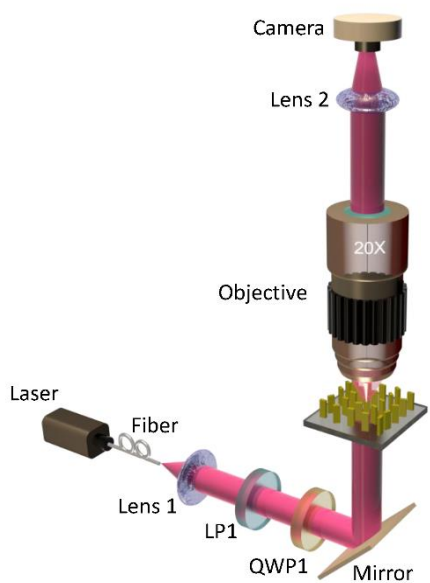

**Supplementary Figure 5.** Experimental setup for measuring the holographic images

The linear polarizer (LP1) and quarter waveplate (QWP1) are employed to guarantee the circular polarization of incident light. To carry out optical performance of our designed hologram under the oblique illumination, the sample is fixed on a mechanical rotator so that we can easily record the tilting angle of sample. The transmitted light is collected directly by an objective lens (Nikon CF plan, 20X, 0.35 NA) and projected onto a charge-couple-device UV camera (JAI, CM-140MCL-UV) by a tube lens. In this system, the objective lens, the tube lens and CCD camera are fixed onto a *z*-axis stage, so that we can tune the distance between objective lens and sample for projecting the holographic image onto the CCD. It should be pointed out that, due to the high conversion efficiency of our meta-hologram, no circular-polarization polarizer is employed for the collection system of holographic image.

The simulated and measured intensity profiles under the oblique illumination are shown in Supplementary Fig. 6. These images could be addressed by the tilting angle ranging from 2° to 32° with an interval of 2°. One can see that the contour, uniformity and sharpness of 'rabbit' pattern get worse when the tilting angle is increased. It is worthy noting that these patterns are still distinguishable at a large angle of 32°, which means an angle of view of 64°×64°. The simulated images are obtained by using the FFT-based Rayleigh-Sommerfeld diffraction, as described in Ref. [S6].

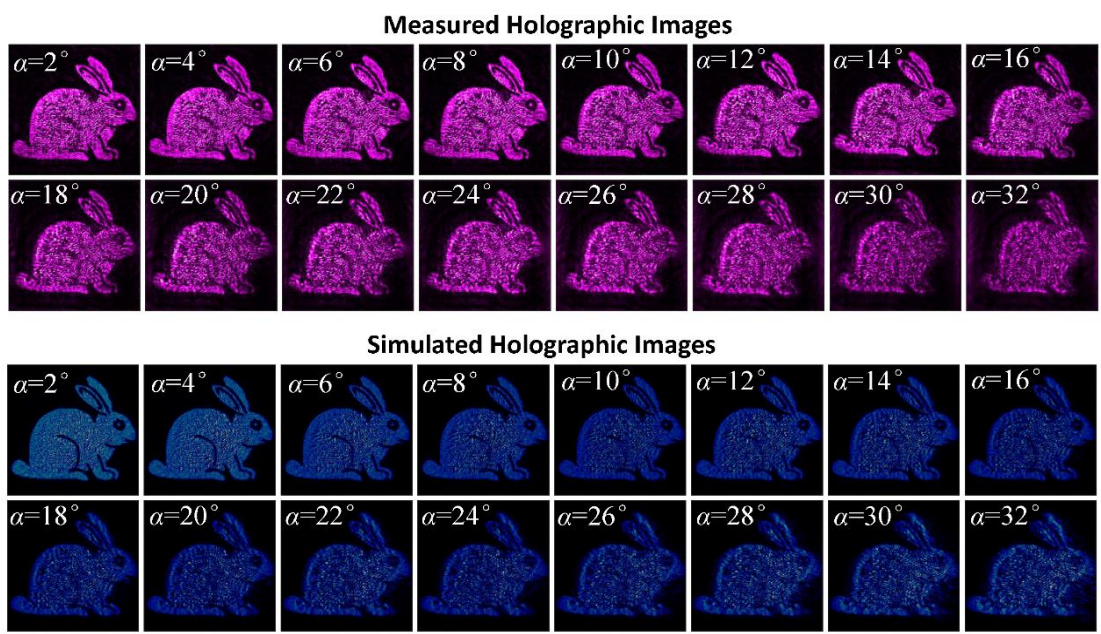

**Supplementary Figure 6.** Measured and simulated holographic images under the oblique illumination with the tilting angle from 2° to 32°.

**Supplementary Note 5 | Measuring the experimental conversion efficiency**

Since our meta-hologram has the imaging plane located in the Fresnel region, the holographic image has the small size and its intensity cannot be measured by using a power meter. To measure the actual conversion efficiency, we record both power of incident light and holographic images by using a CCD camera. The detailed processes are shown in Supplementary Fig. 7.

Firstly, the intensity profile of incident light is recorded by CCD camera. To realize it, we put the sample on the scanning stage of confocal microcopy as shown in Supplementary Fig. 5. By tuning the longitudinal distance between the sample and objective lens, we image the top surface of substrate in the sample and record it as shown in Supplementary Fig. 7a. Note that, it records only the intensity profile of light incident on the top surface of substrate without having any information about hologram.

Secondly, our fabricated hologram with $Nb_2O_5$ nanobricks is moved into the viewing field of microscopy by tuning the *x* and *y* position of the scanning stage (without changing the longitudinal distance between sample and objective lens). Once it is done, the image (see Supplementary Fig. 7b) is captured, where we can find a clear edge (*i.e.*, a white-dashed square of 100μm×100μm) of the $Nb_2O_5$ nanobrick array. Because the incident light is not tuned and has a fixed position in the viewing field of microscopy, both images in Supplementary Figs. 7a and 7b have the same location so that light incident on the white-dashed square can be taken as the total incidence upon the nano-brick array. Thus, in the image of Supplementary Fig. 7a, the power encircled in the white-dahsed square could be taken as the total power $I_{in}$ incident on our metahologram, as shown in Supplementary Fig. 7c.

Finally, we measure the power of holographic images. Keeping the x and y location of scanning stage fixed, we adjust the distance between the sample and objective lens so that we can capture the holographic image of 'rabbit' pattern by using CCD, as shown in Supplementary Fig. 7d. In order to exclude the background noise, we take the intensity encircled in the 'rabbit' region (represented by the red curve) as the output $I_{out}$ of holographic image. The intensity outside the 'rabbit' region is cropped by using Photoshop software. Thus, the experimental efficiency is evaluated by $I_{out}/I_{in}$.

Because our UV camera has high sensitivity, the optical noise from the surrounding environment is measured before recording of each image by CCD. To calculate the experimental efficiency accurately, the environmental noise is extracted from all the images in Supplementary Figs. 7(a) –7(d).

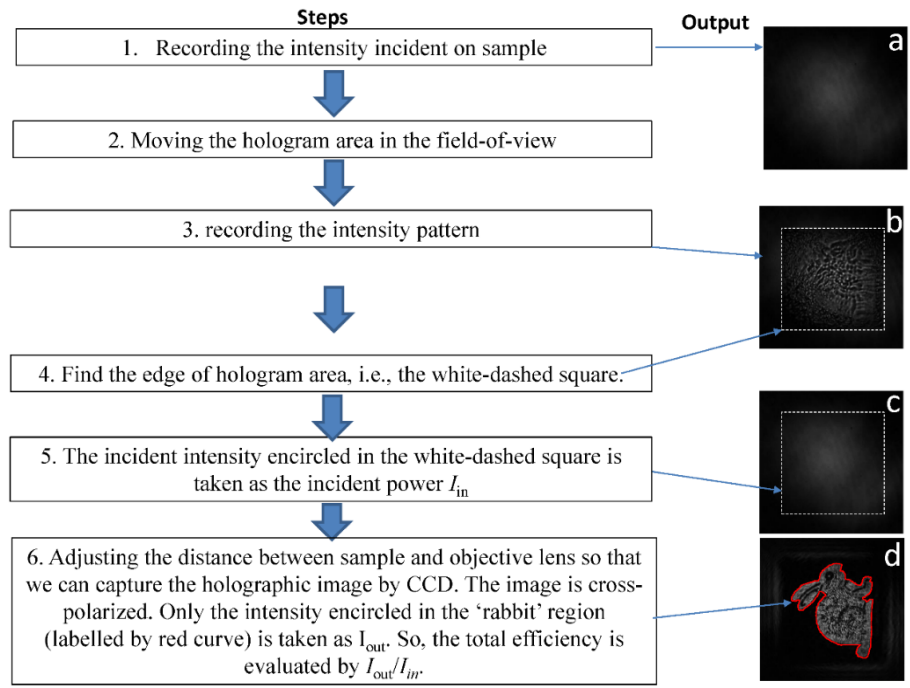

**Supplementary Figure 7.** Detailed steps for measuring the experimental conversion efficiency. Their relative outputs are displayed at the right panel.

**References:**
[S1]. Palik E. D. Ed. Handbook of Optical Constants of Solids. Academic Press: New York, 1991.